\documentclass[preprint,showpacs,preprintnumbers,prd,amssymb]{revtex4}
\RequirePackage[]{amsmath2000}
\let\Box\square

\renewcommand{\theequation}{\arabic{section}.\arabic{equation}}
\def\({\left(}   
\def\){\right)}

\def\lc{\left\{}
\def\rc{\right\}}
\def\t'{t^{\prime}}
\def\r'{r^{\prime}}
\let\Box\square
\newcommand{\be}{\begin{equation}}
\newcommand{\ee}{\end{equation}}
\newcommand{\ben}{$$}
\newcommand{\een}{$$}
\newcommand{\bea}{\begin{eqnarray}}
\newcommand{\eea}{\end{eqnarray}}
\newcommand{\bean}{\begin{eqnarray*}}
\newcommand{\eean}{\end{eqnarray*}}
\newcommand{\e}{{\rm e}}

\newcommand{\R}{\,^4\!R}
\newcommand{\hfa}{high-frequency approximation }
\newcommand{\fourd}{four-dimensional }
\newcommand{\twod}{two-dimensional }
\newcommand{\n}[1]{\label{#1}}
\newcommand{\ind}[1]{\mbox{\tiny{#1}}}

\usepackage{bm}

\begin{document}

\title{
Vacuum polarization in two-dimensional static spacetimes
and dimensional reduction
}
\author{Roberto Balbinot}\email{balbinot@bo.infn.it}
\author{Alessandro Fabbri}\email{fabbria@bo.infn.it}
\author{Piero Nicolini}\email{nicolini@bo.infn.it}
\affiliation{
Dipartimento di Fisica dell'Universit\`a di Bologna and INFN
sezione di Bologna, Via Irnerio 46, 40126 Bologna, Italy}
\author{Patrick J. Sutton}\email{psutton@gravity.phys.psu.edu}
\affiliation{
Center for Gravitational Wave Physics, 
Center for Gravitational Physics and Geometry, 
and Department of Physics, 
The Pennsylvania State University, 
State College, PA, USA 16802-6300
}


\begin{abstract}
We obtain an analytic approximation for the effective action
of a quantum scalar field in a general static \twod
spacetime.  We apply this to the dilaton gravity model
resulting from the spherical reduction of a massive, non-minimally
coupled scalar field in the \fourd Schwarzschild geometry.
Careful analysis near the event horizon shows the resulting \twod system
to be regular in the Hartle-Hawking state for general values of the
field mass, coupling, and angular momentum, while at spatial infinity 
it reduces to a thermal gas at the black-hole temperature. 
\end{abstract}


\maketitle


\section{Introduction}
\label{Introduction}
\setcounter{equation}0

In a spacetime possessing a continuous symmetry one can decompose a quantum
field into harmonics on the symmetrical subspace, effectively reducing the
dimensionality of the system.  For example, a quantum field in a
\fourd spherically symmetric spacetime can be recast as 
a collection of \twod fields labelled by their angular momentum.
If one or a few of these modes make the dominant contribution to
vacuum polarization in four dimensions, solving the reduced
theory only for the modes of interest allows one to truncate
``irrelevant'' degrees of freedom.

This idea motivates much of the literature on \twod black holes.
By restricting attention to the spherically symmetric ``$s$ mode''
of a quantum field, which contributes the bulk of the Hawking
radiation \cite{Ha:75} emitted by Schwarzschild black holes,
several authors \cite{MuWiZe:94,NoOd:97,HaBo:97,BaFa:99,KuLiVa:98}
have proposed \twod effective actions which could be used for the
calculation of back-reaction effects and black-hole evaporation.
These actions have difficulty, however, in reproducing the
expected amount of Hawking radiation, in some cases predicting a
negative flux (anti-evaporation) \cite{BuRa:00}. Before these
actions can be trusted for back-reaction calculations, it is
imperative to understand the behaviour of modes of fixed angular
momentum in the Schwarzschild geometry.

A recent paper \cite{BaFaNiFrSuZe:01} partially addressed this
question.  Using point-splitting regularization and the well-known
WKB approximation, the stress tensor due to the $s$ mode of a
massless, minimally coupled quantum scalar field in the
Schwarzschild geometry was found. A careful analysis demonstrated
that the field is regular at the event horizon in the
Hartle-Hawking state, and also exhibits the expected Hawking flux
at infinity.

In this paper we extend the analysis of \cite{BaFaNiFrSuZe:01} to
general \twod static spacetimes and consider arbitrary harmonics of
massive, nonminimally coupled scalar fields. 
We use the high-frequency approximation \cite{FrSuZe:02,FrSuZe:00,Su:00b} 
to construct an effective action describing the quantum effects of 
a scalar field mode of fixed angular momentum at both zero and 
arbitrary finite temperature.
This action is shown to yield a stress tensor which for the
Schwarzschild spacetime coincides exactly with that obtained in 
canonical quantization through point-splitting regularization 
using the WKB approximation for the modes.  As before 
\cite{BaFaNiFrSuZe:01}, corrections to the WKB and high-frequency 
approximations are necessary to investigate the near-horizon 
behaviour. We show that the stress tensor is regular 
on the Schwarzschild horizon in the Hartle-Hawking state for 
arbitrary mass, coupling and angular momentum, and that it 
reduces to a thermal gas of massive particles at spatial infinity.  

In Section~\ref{dilaton} we review the quantum field--dilaton gravity model.  
Section~\ref{hfa2d} briefly outlines the high-frequency approximation 
and applies it to our system, producing analytic expressions for the 
effective action and stress tensor throughout the spacetime. Special 
attention is devoted to the region near the horizon in 
Section~\ref{fix}, where we demonstrate the regularity
of the field in the Hartle-Hawking state. We conclude with some
brief comments in section~\ref{conclusions}.

\section{Quantum Field Coupled to Dilaton Gravity}
\label{dilaton}
\setcounter{equation}0

We wish to construct the \twod theory corresponding to a mode of
fixed angular momentum of a scalar field in a four-dimensional
static \footnote{A static geometry is a requirement of our approximation
scheme.}
spherically symmetric geometry.
The line element may be written as
\be\n{4dmetric}
ds^2  =  h_{ab}(x^c) dx^a dx^b
         +\rho^2\e^{-2\phi(x^c)} \Omega_{ij}(y^k) dy^i dy^j \, ,
\ee
where $h_{ab}$ is a static two-dimensional Euclidean metric,
$\Omega_{ij}$ is the metric of the unit two-sphere,
$\rho$ is a constant with dimensions of length, and
$r\equiv\rho\e^{-\phi}$; the function $\phi$ is termed the {\em dilaton}.
The \fourd scalar field $\widehat{\Phi}$ satisfies the equation
\be\n{4dfieldeqn}
(\Box -m^2 -\xi \R) \, \widehat{\Phi} = 0 \, ,
\ee
where $\Box$ and $\R$ are the d'Alembertian and scalar curvature in
four dimensions.

Our model is obtained by decomposing $\widehat{\Phi}$
in terms of spherical harmonics $Y_{lm}(y^i)$ on the two-sphere:
\be\n{decomposition}
\widehat{\Phi}(x^c,y^k)
  =  \sum_{l=0}^\infty \sum_{m=-l}^l \, \widehat{\varphi}_l(x^c) \,
     \frac{Y_{lm}(y^i)}{r} \, .
\ee
The \twod field $\widehat{\varphi}_l$ satisfies
\be\n{2dfieldeqn}
(\Delta - m^2 - V) \, \widehat{\varphi}_l = 0 \, ,
\ee
where $\Delta$ is the d'Alembertian for the metric $h_{ab}$,
the potential $V$ is
\be\n{2dpotential}
V  =  \xi R + \frac{1}{r^2}[2\xi+l(l+1)]
      + (4\xi-1)\Delta\phi + (1-6\xi)(\nabla\phi)^2 \, ,
\ee and $R$ is the \twod scalar curvature associated with
$h_{ab}$. Equations (\ref{2dfieldeqn})--(\ref{2dpotential}) define
the 2D quantum field theory which we will study.  
The special case $m=0$, $\xi=0$, $l=0$ in a Schwarzschild geometry
was dealt with in \cite{BaFaNiFrSuZe:01} (see also
\cite{MuWiZe:94,NoOd:97,BaFa:99,KuLiVa:98}).

In Section~\ref{hfa2d} we use the \hfa to calculate the one-loop effective
action $W$ for (\ref{2dfieldeqn})--(\ref{2dpotential}).
Given $W$, the expectation value of the stress tensor can be obtained as
\be\n{stressW}
\langle T_{ab} \rangle
  =  \frac{2}{\sqrt{h}} \frac{\delta W}{\delta h^{ab}\vphantom{\sqrt{h}}} \, .
\ee 
This \twod stress tensor is {\em not} conserved \cite{BaFa:99,Su:00b}; 
rather, it satisfies 
\be\n{conservation} \nabla_a
\langle T^{ab} \rangle
  =  -2\nabla^b\phi \, P \, ,
\ee
where the ``effective pressure'' $P$ corresponds
to $\langle T_\theta{}^\theta \rangle$ from four dimensions.
This pressure is also easily calculated from the effective action:
\be\n{pressureW}
P  =  \frac{1}{2\sqrt{h}} \frac{\delta W}{\delta \phi} \, .
\ee
Equations (\ref{conservation}), (\ref{pressureW}) follow from
the conservation of the full \fourd stress tensor.  
Using the relationship (\ref{decomposition}) between the four-dimensional 
field $\widehat{\Phi}$ and its two-dimensional counterpart 
$\widehat{\varphi}_l$, one can express the four-dimensional stress tensor 
in terms of the two-dimensional one.  Substitution into the 
four-dimensional conservation equation then yields (\ref{conservation}).  
For details, see \cite{BaFa:99,Su:00a,Su:00b}.

Local observables may also be calculated using the Euclidean 
Green function, which satisfies 
\be\n{fieldeqn} 
(\Delta-m^2-V)\,G(x,x) = -\frac{\delta(x-x')}{\sqrt{h}} \,  
\ee 
and is related to the effective action by the formal expression 
\be\n{trlnG}
W = -\frac{1}{2}\,\mathrm{Tr}\,\ln{G} \, .
\ee
The stress tensor and pressure can be obtained by applying 
certain differential operators to the Green function:
\begin{subequations}\n{stress}
\be\n{stressG}
\langle T_{ab} \rangle = \lim_{x'\to x} \, D_{ab} \, G(x,x') \, ,
\ee
\be\n{pressureG}
P = \lim_{x'\to x} \, D \, G(x,x') \, ,
\ee
where \cite{Su:00b}
\bea\n{stressD}
D_{ab}
  & \equiv &
         (\frac{1}{2}-\xi) \left(
             h_a{}^{c'} \nabla_{c'}\nabla_b
             +h_b{}^{c'} \nabla_a\nabla_{c'}
         \right)
         -\xi\left(
             \nabla_a\nabla_b
             +h_a{}^{c'} h_b{}^{d'} \nabla_{c'}\nabla_{d'}
         \right)
         \nonumber \\
  &   &  \mbox{}
         +(2\xi-\frac{1}{2})h_{ab}
             h^{cd'}\nabla_c\nabla_{d'}
         +\xi R_{ab}
         +(2\xi-\frac{1}{2})h_{ab}\phi^c\left(
             \nabla_c + h^{d'}_c \nabla_{d'}
         \right)
         \nonumber \\
  &   &  \mbox{}
         -(2\xi-\frac{1}{2})\left[
             \phi_a(\nabla_b + h^{d'}_b \nabla_{d'})
             +\phi_b(\nabla_a + h^{d'}_a \nabla_{d'})
         \right]
         +(1-6\xi)\phi_a\phi_b
         \nonumber \\
  &   &  \mbox{}
         +(2\xi-\frac{1}{2})h_{ab}\left[
             m^2 + \xi R
             +4\xi\Delta\phi+(1-6\xi)(\nabla\phi)^2
             +\frac{1}{r^2}(2\xi + l(l+1))
         \right] \, , \qquad 
\eea
\bea\n{pressureD}
D
  & \equiv &
         (2\xi-\frac{1}{2}) h^{cd'}\nabla_c\nabla_{d'}
         -\frac{1}{2}\,(1-6\xi)\left[
             \phi^c\left(\nabla_c + h^{d'}_c \nabla_{d'}\right)
             +\Delta\phi
         \right]
         \nonumber \\
  &   &  \mbox{}
         +(2\xi-\frac{1}{2})\left[
             m^2 + \xi R
             +(4\xi-1)\Delta\phi
	     +(1-6\xi)(\nabla\phi)^2
	 \right] 
	 +\frac{2\xi}{r^2}\,(2\xi + l(l+1))
         \, . \qquad \quad
\eea
\end{subequations}
Here $h_a{}^{b'}$ is the bivector of parallel transport, and
$\phi_a\equiv\phi_{;a}$.
This technique will be used to calculate corrections to the \hfa
near the event horizon in Section~\ref{fix}.

\section{High-Frequency Approximation in Dilaton Gravity}
\label{hfa2d}
\setcounter{equation}0

The \hfa \cite{FrSuZe:02,FrSuZe:00,Su:00b} is a new generalized
approximation scheme for quantum fields in static spacetimes. 
Here we apply this scheme to our dilaton gravity model
(\ref{2dfieldeqn})--(\ref{2dpotential}). In
Appendix~\ref{WKBappendix} we show that for the $s$ mode ($l=0$)
in the reduced Schwarzschild spacetime our \hfa is equivalent to
the WKB approximation; however, the \hfa is simpler and gives the
effective action for general static spacetimes and 
field modes. Our analysis will be
brief; for additional details see \cite{FrSuZe:00,FrSuZe:02,Su:00b}.

\subsection{High-Frequency Approximation in Two Dimensions}

The \hfa models the observable of interest, in this case the 
effective action, using the ultraviolet divergences of the theory.  
Since high frequencies are equivalent to short distances, 
one is in essence assuming that the dominant features of the field 
(aside from any temperature dependence) 
can be approximated using its local behaviour.   

The procedure is simple: 
\begin{enumerate}
\item 
Use as the starting point the Schwinger-DeWitt short-distance expansion 
\cite{Ch:76,BiDa:82} for the observable of interest, which is a local 
expression containing all of the ultraviolet divergences of the theory.
Since only this ultraviolet behaviour is of interest, discard any terms 
in the expansion which are finite in the coincidence limit.
\item
Fourier transform over the time splitting, expand in a Laurent
series in the inverse frequency $\omega^{-1}$ as
$\omega\to\infty$, and retain only those terms
which diverge when summed over $\omega$. 
(These are the only terms which are unambiguously specified by the
ultraviolet divergences of the theory.) 
This defines the \hfa in the frequency domain.
\item
Return to the time domain by summing over all frequencies 
(performing the inverse Fourier transform).
\item
Renormalize in the usual way for point splitting, by subtracting the 
Schwinger-DeWitt expansion.
\end{enumerate}

Let us apply this scheme to our dilaton-gravity model. 
We may write the line element of a two-dimensional static spacetime as
\be\label{line}
ds^2  =  h_{ab}dx^a dx^b
      =  \e^{-4\eta(x)}dt^2 + dx^2 \, ,
\ee
where $\eta$ is a function of the coordinate $x$ but not of $t$.
Then for infinitesimally separated points, the Schwinger-DeWitt 
expansion \cite{Ch:76,BiDa:82} for the Lagrangian of the effective 
action (\ref{trlnG}) is  
\bea\n{LDS}
L_{\ind{DS}}(t,x;t',x')
  & = &  \mbox{}
         - \frac{1}{2\pi(\tau^2+\epsilon^2)}
         +\frac{1}{8\pi} \left(\frac{1}{6} R-V-m^2\right)
             \ln\frac{m^2(\tau^2+\epsilon^2)}{4\e^{-2\gamma}}
         \nonumber \\
  &   &  \mbox{}
         +\frac{m^2}{8\pi}
         -\frac{1}{12\pi}\Delta\eta
         -\frac{1}{6\pi}(\nabla\eta)^2
     + {\cal O}(\tau^2,\epsilon^2) \, ,
\eea
where $\tau^2 = \e^{-4\eta}(t-t')^2$, $\epsilon = (x-x')^2$,
and $\gamma$ is the Euler-Mascheroni constant.

To construct our approximation, we Fourier transform over the 
time splitting:
\be\n{FT}
L(x;x'|\omega)
  =  \int_{-\infty}^{\infty} \!\!dt\,
     \e^{i\omega t} L(t,x;0,x') \, .
\ee
Since we are only interested in the high-frequency limit of
$L(x;x'|\omega)$, we need accurately model only the divergences
of $L(t,x;t',x')$.  Casting off terms in (\ref{LDS}) which are finite
in the coincidence limit \footnote{There is an ambiguity in how 
one specifies the divergent part of the logarithm term in (\ref{LDS}).
Our choice corresponds to using 
$\displaystyle{ \ln{\frac{m^2(\tau^2+\epsilon^2)}{4\e^{-2\gamma}}}
  \to  \ln{\frac{m^2(\tau^2+\epsilon^2)}{1+m^2\tau^2}}  }$.}
and using equations (3.723.2) and (4.382.3)
of \cite{GrRy:94}, we arrive at \footnote{
$L(x,x'|\omega)$ also contains a $1/\epsilon$ term which is the usual
leading divergence for a one-dimensional action; it is the `superficial'
divergence encountered by Candelas and Howard \cite{CaHo:84} and others
\cite{An:90,AnHiSa:95}, and may be safely ignored.}
\be\n{LwHFA}
L_{\ind{HFA}}^{\ind{bare}}(x;x|\omega)  \equiv
    \frac{|\omega|}{2\chi^2}
    -\frac{1}{4|\omega|} \left(\frac{1}{6} R-V-m^2\right) \, ,
\ee
where $\chi^2 = \e^{-4\eta}$ is the norm of the Killing vector
$\chi^a$ for the static space (\ref{line}), and we have dropped
all terms of ${\cal O}(\omega^{-3})$ and higher.

We return to the time domain by summing (\ref{LwHFA}) over 
all frequencies.  For a scalar field at finite temperature $T=\beta^{-1}$
the allowed frequencies are discrete,
\be
\omega_n = n\,\frac{2\pi}{\beta} \, , \n{frequencies}
\ee
where $n$ is an integer.  (From (\ref{frequencies}), it is clear that the 
\hfa includes an assumption of high temperature.)  
The bare approximate effective Lagrangian in the time domain can then
be evaluated using equation (E1) of \cite{AnHiSa:95}, giving
\bea
L_{\ind{HFA}}^{\ind{bare}}(t,x;t',x)
  & = &  \frac{2}{\beta} \,
         \sum_{n=1}^{\infty} \, \cos(\omega_n (t-t')) \,
	 L_{\ind{HFA}}^{\ind{bare}}(x;x|\omega_n)
         \n{Lsum} \\
  & = &  \mbox{}
         -\frac{1}{2\pi\tau^2}-\frac{\pi}{6\chi^2\beta^2}
         +\frac{1}{8\pi}\left(\frac{1}{6} R-V-m^2\right)
         \ln\frac{4\pi^2\tau^2}{\chi^2\beta^2} \, . \n{LHFA}
\eea
Note that the zero-frequency mode makes no contribution in
the high-frequency approximation; from (\ref{LwHFA})--(\ref{frequencies}), 
it consists of one term proportional to $n$, which vanishes, and one
term proportional to $1/n$, which must be dropped to regularize
the infrared behaviour of the Lagrangian.

The ultraviolet divergences are renormalized in the usual way
by subtracting the Schwinger-DeWitt expansion (\ref{LDS}); integrating
over the spacetime then gives the renormalized effective action
\bea\label{WHFA}
W^{\ind{ren}}_{\ind{HFA}}
  & = &  \int \!d^2x \sqrt{h}\, \left[ \mbox{}
             -\frac{\pi}{6\chi^2\beta^2}
             -\frac{m^2}{8\pi}
             +\frac{(\nabla\eta)^2}{6\pi}
             -\frac{1}{8\pi}\ln{\left(
                 \frac{m^2\beta^2\chi^2}{16\pi^2\e^{-2\gamma}}
             \right)} \left(\frac{1}{6} R - V - m^2 \right)
         \right]   , \qquad   
\eea
where we have discarded a surface term.  This action can be put into
explicitly covariant form by the substitution
\be\n{eta}
\eta = -\frac{1}{4}\ln{\chi^2} \, .
\ee
All further expectation values will be calculated 
from $W^{\ind{ren}}_{\ind{HFA}}$ and hence will be renormalized already, 
so the superscript `ren' will be dropped.

To determine the stress tensor we must specify the
potential $V$.  Using (\ref{2dpotential}),
(\ref{stressW}), and (\ref{pressureW}), we find
\bea
\langle T_{ab} \rangle_{\ind{HFA}}
  & = &  \frac{\pi}{6\beta^2\chi^2}\,\left[
             h_{ab} - 2\frac{\chi_a\chi_b}{\chi^2}
         \right]
         +\frac{m^2}{8\pi}\,h_{ab} - \frac{1}{6\pi}h_{ab}(\nabla\eta)^2
         +\frac{1}{3\pi}\nabla_{\!\!a}\eta\nabla_{\!\!b}\eta
         -\frac{1}{6\pi}\frac{\chi_a\chi_b}{\chi^2}\Delta\eta
         \nonumber \\
  &   &  \mbox{}
         +\frac{1}{4\pi}\frac{\chi_a\chi_b}{\chi^2}\left[ \,
             (\frac{1}{6}-\xi) R
             - m^2
             - \frac{1}{r^2}[2\xi+l(l+1)]
             + (1-4\xi)\Delta\phi
             + (6\xi-1)(\nabla\phi)^2 \,
         \right]
         \nonumber \\
  &   &  \mbox{}
         -\frac{1}{2\pi}\left[ \,
             (\frac{1}{6}-\xi)2[\nabla_{\!\!a}\nabla_{\!\!b}\eta
                 - h_{ab}\Delta\eta]
             +(1-4\xi)[\nabla_{\!\!a}\phi\nabla_{\!\!b}\eta
                 +\nabla_{\!\!b}\phi\nabla_{\!\!a}\eta
                 -h_{ab}\nabla\phi\!\cdot\!\nabla\eta] \,
         \right]
         \nonumber \\
  &   &  \mbox{}
         +\frac{1}{8\pi}\ln{\left(
         \frac{m^2\beta^2\chi^2}{16\pi^2\e^{-2\gamma}}
     \right)}
         \left[ \,
             h_{ab}\left(
                 - m^2
                 - \frac{1}{r^2}[2\xi+l(l+1)]
                 + (6\xi-1)(\nabla\phi)^2
             \right)
             \right. \nonumber \\
  &   &      \hspace{0.0in} \left. \vphantom{\frac{1}{1}}\mbox{}
             +2(1-6\xi)\nabla_{\!\!a}\phi\nabla_{\!\!b}\phi \,
         \right] \, ,  \n{stressdil} \\
P_{\ind{HFA}}
  & = &  \frac{1}{8\pi}\left[ \,
             \vphantom{\sum}
             (1-4\xi) 2 \Delta\eta
             +(1-6\xi) 4 \nabla\phi\!\cdot\!\nabla\eta
         \, \right] \nonumber \\
  &   &  \mbox{}
         +\frac{1}{8\pi}\ln{\left(
         \frac{m^2\beta^2\chi^2}{16\pi^2\e^{-2\gamma}}
     \right)}
         \left[ \,
             (6\xi-1)\Delta\phi
             +\frac{1}{r^2}[2\xi+l(l+1)] \,
         \right]  \, ,
         \n{pressuredil}
\eea
where use of 
\be 
\frac{\delta\eta}{\delta h^{ab}} =
\frac{\chi_a\chi_b}{4\chi^2}
\ee 
has been made.
One can verify that the stress tensor and pressure obey the
``conservation'' equation (\ref{conservation}). For the
conformally invariant theory ($m=0$, $\xi=0$, and $l=0$), this
stress tensor also has the correct anomalous trace \cite{BiDa:82} of 
\be
\langle T_\nu{}^\nu \rangle_{\ind{HFA}} =
\frac{1}{4\pi}\left(\,\frac{1}{6}\,R-V\,\right) \, .
\ee

Equations (\ref{stressdil})--(\ref{pressuredil}) provide analytic 
expressions for expectation values for a wide class of 
quantum scalar fields in general static \twod spacetimes. For 
example, for the Polyakov theory ($m=0$, $\xi=0$, $l=0$, $\phi=0$) 
the effective action (\ref{WHFA}) reduces to the exact result for 
a static geometry \cite{FrIsSo:96}.  Unfortunately, for more 
general systems these expectation values all contain $\ln{\chi^2}$, 
and thus typically will diverge at horizons, even in the 
Hartle-Hawking state.  This appears to be a generic feature of local 
approximation schemes, occuring also in the WKB \cite{AnHiSa:95,An:90},
Page-Brown-Ottewill \cite{Pa:82}, and Killing \cite{FrZe:87}
approximations in both two and four dimensions.
In Section~\ref{fix} we shall see how the \hfa
can be corrected near the horizon through more accurate modelling of
the lowest-frequency field modes.

\subsection{Spherically Reduced Schwarzschild Geometry}
\label{srg}

For the special case of a field propagating in the Schwarzschild geometry,
\be\n{metric}
ds^2 = f(r)\, dt^2 + \frac{1}{f(r)}\, dr^2 \,
+ r^2 (d\theta^2+\sin^2{\theta}d\phi^2),
\qquad f(r) = 1-\frac{2M}{r} \, ,
\ee
we have
\be
\chi^2 = f \, , \qquad 
\eta = -\frac{1}{4}\ln{f} \, , \qquad 
\phi = -\ln{\frac{r}{\rho}} \, ,
\ee
where $\rho$ is some arbitrary length scale.   
The \hfa for the stress tensor and pressure for
general $m$, $\xi$ and $l$ is
\begin{subequations}\label{Schw}
\bea
\langle T_t{}^t \rangle_{\ind{HFA}}
  & = &  -\frac{\pi}{6\beta^2f}
         +\frac{1}{24\pi r^2f}\left[\,
             -2\frac{M}{r}+5\frac{M^2}{r^2}
             +12\xi\left(-4\frac{M}{r}+7\frac{M^2}{r^2}\right)
         \,\right]
         -\frac{m^2}{8\pi}
         -\frac{l(l+1)}{4\pi r^2}
         \nonumber \\[1mm]
  &   &  \mbox{}
         +\frac{1}{8\pi r^2}\ln{\left(
             \frac{m^2\beta^2f}{16\pi^2\e^{-2\gamma}}
     \right)}
         \left[
             -\left(1-\frac{2M}{r}\right)
             +4\xi\left(1-\frac{3M}{r}\right)
             -m^2r^2
             -l(l+1)
         \right]
         \, , \nonumber \\
  &   &  \label{SchwTtt} \\[0mm]
\langle T_r{}^r \rangle_{\ind{HFA}}
  & = &  \frac{\pi}{6\beta^2f}
         +\frac{1}{24\pi r^2f}\left[\,
             -6\frac{M}{r}+11\frac{M^2}{r^2}
             +12\xi\left(2\frac{M}{r}-3\frac{M^2}{r^2}\right)
         \,\right]
         +\frac{m^2}{8\pi}
         \nonumber \\[1mm]
  &   &  \mbox{}
         +\frac{1}{8\pi r^2}\ln{\left(
             \frac{m^2\beta^2f}{16\pi^2\e^{-2\gamma}}
     \right)}
         \left[
             \left(1-\frac{2M}{r}\right)
             +4\xi\left(-2+\frac{3M}{r}\right)
             -m^2r^2
             -l(l+1)
         \right]
         \, , \nonumber \\
  &   &  \label{SchwTrr} \\[0mm]
P_{\ind{HFA}}
  & = &  \frac{1}{8\pi r^2}\left[ \frac{4M}{r} - \xi \frac{20M}{r} \right]
         \nonumber \\[1mm]
  &   &  \mbox{}
         +\frac{1}{8\pi r^2}\ln{\left(
             \frac{m^2\beta^2f}{16\pi^2\e^{-2\gamma}}
     \right)}
         \left[
             -1 + \frac{4M}{r} + 8\xi\left(1-3\frac{M}{r}\right) + l(l+1)
         \right]
         \, . \label{SchwP}
\eea
\end{subequations}
For $l=0$ and general $m$ and $\xi$, these expectation values 
coincide with the WKB approximation, as shown in 
Appendix~\ref{WKBappendix}.
For a zero-temperature conformally invariant field
($\beta\to\infty$, $m=0$, $\xi=l=0$, with the $m$ in the
logarithm terms replaced by an arbitrary cutoff parameter), 
(\ref{Schw}) also coincides with the stress tensor and pressure 
obtained from integrating the \twod conformal anomaly 
alone \cite{BaFa:99}. 

At spatial infinity, the stress tensor (\ref{SchwTtt}), (\ref{SchwTrr}) 
reduces precisely to that of a thermal gas of massive particles at 
high temperature in flat space, except that 
$\langle T_r{}^r\rangle$ is missing a term proportional to $m/\beta$. 
This term is lost because the high-frequency approximation does not 
include the contribution of the zero-frequency mode, which must be 
added by hand.  Using the WKB approximation presented in 
Appendix~\ref{WKBappendix}, one finds that at spatial infinity the only 
nonvanishing contribution of the $n=0$ mode is 
\be
\langle T_r{}^r \rangle_{n=0|WKB} = -\frac{m}{2\beta\sqrt{f}}\, .
\ee
Adding this term to (\ref{SchwTrr}) reproduces exactly the stress tensor 
of a thermal gas of massive particles at high temperature in flat space.
This stress tensor obeys the conservation equation (\ref{conservation}), 
with the right-hand side of (\ref{conservation}) vanishing at spatial 
infinity.

The situation at the horizon is more troublesome.  
It was seen in \cite{BaFaNiFrSuZe:01} that for the 
conformally invariant field in the Hartle-Hawking
state ($\beta=8\pi M$) the stress tensor and
pressure (\ref{Schw}) diverge logarithmically at the 
event horizon in a freely falling frame \footnote{
The stress tensor will be finite
in a freely-falling frame on the past and future event horizons
provided that as $r\rightarrow2M$
\ben
\left|T_t{}^t + T_r{}^r\right|  <  \infty
\, , \qquad
f^{-1}\left|T_t{}^t - T_r{}^r\right|  <  \infty
\, , \qquad
\left|P\right|  <  \infty
\, .
\een
}
due to the logarithm terms.  For nonconformal fields 
(any of $m$, $\xi$, or $l$ nonzero), the additional terms in 
(\ref{Schw}) produce yet stronger divergences on the past and 
future horizons.  Since the Hartle-Hawking state should be 
regular on the horizon, this implies that the high-frequency 
and WKB approximations break down there.  Some indication of 
why comes from inspection of (\ref{LDS}), (\ref{LHFA}): the 
logarithm term in the bare Lagrangian (\ref{LHFA}) 
does not contain the norm of the Killing vector [note that 
$\tau^2/\chi^2=(t-t')^2$]; the $\ln{\chi^2}$ terms arise purely from 
the renormalization counterterms (\ref{LDS}).  This implies
that the bare \hfa misses contributions which 
go as $\ln{\chi^2}$ near the horizon.  In the following section 
we demonstrate that this omission occurs because the \hfa
fails to accurately model the lowest-frequency modes ($n\le2$) near the
horizon.  Correcting for these modes shows that the stress tensor 
and pressure are indeed regular on the horizon in the Hartle-Hawking state.


\section{Regularity of the Hartle-Hawking State}
\label{fix}
\setcounter{equation}0

In this section we demonstrate the regularity of the stress tensor
and pressure at the event horizon of the Schwarzschild geometry
in the Hartle-Hawking state for general $m$, $\xi$, and $l$.
We do this by solving the Green function equation (\ref{fieldeqn})
near the horizon for individual modes $\omega_n=2\pi n/\beta$ of small $n$,  
and using these more accurate expansions to correct the high-frequency 
approximation.  This analysis closely follows that of \cite{BaFaNiFrSuZe:01}.

\subsection{Mode-by-Mode Predictions of the High-Frequency Approximation}

We begin by listing the contributions made by each frequency $n$ to
expectation values in the high-frequency approximation.  This will be 
necessary later for determining the amount by which the approximation
must be corrected for each mode.

Using (\ref{LwHFA})--(\ref{Lsum}) one finds that for each $n>0$ the bare 
contributions to the stress tensor and pressure for $\beta=8\pi M$ are
\begin{subequations}\label{HFAmode}
\bea
\langle T_a^b \rangle_{n|\ind{HFA}} 
  & = &  \frac{n}{32\pi M^2f}\left[ 2\delta_a^0\delta_0^b-\delta_a^b\right] 
         \nonumber \\
  &   &  \mbox{}+\frac{1}{4\pi n}\left[
	     -\delta_a^b(m^2+\frac{\ell(\ell+1)}{r^2}+\frac{2\xi}{r^2})
	     +(1-6\xi)\frac{f}{r^2}(2\delta_a^r\delta_r^b-\delta_a^b)
	 \right]  \, , \n{HFAT} \\
P_{n|\ind{HFA}} 
  & = &  \frac{1}{4\pi r^2 n}\left[
             (6\xi-1)(1-\frac{4M}{r})+2\xi+\ell(\ell+1)
         \right] \, . 
\eea
\end{subequations}
Note the absence of any logarithmic terms.
Also, recall that the \hfa includes no contributions from the $n=0$ mode.

\subsection{Corrected Green Function near the Horizon}

We now proceed to calculate the exact contribution to the Green 
function and local observables made by each mode $n$ in the 
vicinity of the event horizon of the Schwarzschild geometry. 
We do this by solving for the Green function as a 
power series in the physical distance to the event horizon. 
We assume the field to be in the Hartle-Hawking state 
($\beta=8\pi M$).

In the $(t,r)$ sector of the Schwarzschild spacetime (\ref{metric})
we decompose the Green function (\ref{fieldeqn}) into Fourier modes as
\be\label{Gdecomp}
G(t,r;t',r')
  =  \frac{1}{\beta[f(r)f(r')]^{\frac{1}{4}}} \left[
         G_0(r,r') + 2 \sum_{n=1}^{\infty}
         \cos{(\omega_n (t-t'))}\, G_n(r,r')
     \right] \, .
\ee
Substituting (\ref{Gdecomp}) and a similar decomposition for
the delta function into the differential equation (\ref{fieldeqn})
for the two-dimensional Green function, and expanding in terms of
the physical distance $L$ to the event horizon, where
\be
dL = \frac{dr}{f^{\frac{1}{2}}} \, ,
\ee
the differential equation with $r\ne r'$ becomes
\be\label{ODE}
\partial_L^2G_n
-\left(\,
    \frac{4n^2-1}{4L^2}+\frac{\alpha^2_n}{M^2}-\frac{A^2_n}{16M^4}L^2+O(L^4)\,
\right)\,G_n
  = 0 \, ,
\ee
where we define the constants
\be
\alpha^2_n \equiv (mM)^2 + \frac{l(l+1)}{4}
  + \frac{1}{6} + \frac{n^2}{12} \, ,
\ee
\be
A^2_n \equiv \frac{n^2 + 56 + 60l(l+1)}{120} \, .
\ee
Neglecting the $O(L^2)$ term, (\ref{ODE})
becomes a differential equation for the Bessel functions of imaginary
argument.  The solution for $G_n$ is easily shown to be
\be\label{soln}
G_n(r,r') = (LL')^{\frac{1}{2}} I_n\left(\frac{\alpha_n L_<}{M}\right)
                   K_n\left(\frac{\alpha_n L_>}{M}\right) \, ,
\ee
where $L_>$ ($L_<$) is the physical distance to the greater (lesser) of
$r$, $r'$.  This solution obeys the derivative condition resulting from
integrating across the delta-function singularity at $r=r'$.

The approximate solution (\ref{soln}) will yield the pressure
accurately to $O(1+\ln{L})=O(1+\ln{f})$ inclusive, which contains
all of the potentially divergent terms.  To prove the finiteness
of the stress tensor, however, it is also necessary to calculate
the $O(L^{\frac{1}{2}}\ln{L})=O(f\ln{f})$ terms; a careful
examination reveals that this requires taking the $O(L^2)=O(f)$
term in (\ref{ODE}) into account.  The simplest way to do this is
to begin with the approximate solution (\ref{soln}), 
expand it for small $L$ and $L'$, and
modify the higher-order terms so that the resulting Green function
satisfies (\ref{ODE}) up to and including the $O(L^2)$ term.
The resulting Green functions $G_n$ for $n=0,1,2,3$ are
listed in Appendix~\ref{correctedG} .

\subsection{Corrections to Local Observables}

We now calculate the precise contributions to the stress
tensor and pressure for modes of small $n$, and use this information
to correct the high-frequency approximation.

Using (\ref{stress}), for a contribution to the \twod 
Green function of the form
\be
\cos{(\omega_n(t-t'))} \,F_n(r,r')  \, ,
\ee
the corresponding contributions to the stress tensor and pressure
in $t,r$ coordinates are
\begin{subequations}
\bea
\langle T_t^t \rangle_n
  & = &  \lim_{r'\rightarrow r}
         \left[
         (2\xi+\frac{1}{2})\frac{\omega_n^2}{f}
         +\xi \frac{2M}{r^3}
         +(2\xi-\frac{1}{2})[m^2+\frac{f+l(l+1)}{r^2}]
         \right.\nonumber \\
  &   &      \hspace{1cm} \mbox{} \left.
             +[\frac{1}{2}f+\xi(-2+\frac{3M}{r})]
             \frac{1}{r}(\partial_r+\partial_{r'})
         +(2\xi-\frac{1}{2}) f \partial_r \partial_{r'}
     \right] F_n(r,r')
         \, , \\ 
\langle T_r^r \rangle_n
  & = &  \lim_{r'\rightarrow r}
         \left[
         (2\xi-\frac{1}{2})\frac{\omega_n^2}{f}
         +\xi \frac{2M}{r^3}
         +(2\xi-\frac{1}{2})[m^2+\frac{f+l(l+1)}{r^2}]
         +(1-6\xi)\frac{f}{r^2}
         \right.\nonumber \\
  &   &      \hspace{1cm} \mbox{} \left.
             +[-\frac{1}{2}f+\xi(2-\frac{5M}{r})]
             \frac{1}{r}(\partial_r+\partial_{r'})
         +\frac{1}{2} f \partial_r \partial_{r'}
         -\xi f (\partial_r^2+\partial_{r'}^2)
     \right] F_n(r,r')
         \, ,  \qquad \quad \\ 
P_n
  & = &  \lim_{r'\rightarrow r}
         \left[
         (2\xi-\frac{1}{2})\frac{\omega_n^2}{f}
         +2\xi\frac{f}{r^2}+\frac{l(l+1)}{2r^2}
         +(2\xi-\frac{1}{2})[m^2+\frac{f+l(l+1)}{r^2}]
         \right.\nonumber \\
  &   &      \hspace{1cm} \mbox{} \left.
             +(1-6\xi)\frac{f}{2r}(\partial_r+\partial_{r'})
         +(2\xi-\frac{1}{2}) f \partial_r \partial_{r'}
     \right] F_n(r,r')
         \, . 
\eea
\end{subequations}
Using the $G_n$ in Appendix~\ref{correctedG} one can calculate 
the contribution to the stress tensor from each mode near the horizon.  
Subtracting the corresponding \hfa result (\ref{HFAmode}) yields the 
corrections that should be added to the stress tensor and pressure for 
each $n$.  Doing so, one finds that divergent corrections come only 
from the $n=0,1,2$ modes; higher modes give only corrections which are 
regular at the horizon.  The total divergent corrections to be added 
to the \hfa are
\begin{subequations}
\bea
\delta\langle T_a{}^b \rangle_{\ind{div~corr}}
  & = &  \frac{1}{96\pi M^2} \, \left[ \,
             \frac{3\xi}{f}
             +(12\,\mu^2 + 6\,\xi + 3\, l(l+1))
             +(3 - 18\,\xi) f \ln{f}
         \, \right] \left(
             \begin{array}{r r} 1 & 0 \\ 0 & -1 \end{array}
         \right)
         \nonumber \\
  &   &  \mbox{}
         +\frac{1}{96\,\pi\,M^2}\,(12\,\mu^2 + 6\,\xi  + 3\, l(l+1))\,\ln{f}
         \left(
             \begin{array}{r r} 1 & 0 \\ 0 & 1 \end{array}
         \right) \, , \\
\delta P_{\ind{div~corr}}
  & = &  \frac{1}{96\,\pi\,M^2}\,(-3 + 12\,\xi - 3\, l(l+1))\,\ln{f}\, .
\eea 
\end{subequations}
For the conformal field these reproduce the divergent terms 
found in \cite{BaFaNiFrSuZe:01} (note that in \cite{BaFaNiFrSuZe:01} 
$P$ was defined with an additional factor of $(4\pi r^2)^{-1}$).  
By adding these corrections to the \hfa
(\ref{Schw}) one finds that the resulting stress
tensor and pressure are indeed regular on the event horizon for
general $m$, $\xi$, and $l$.

The corrected stress tensor and pressure near the horizon,  
including all finite corrections for the $n=0,1,2$ modes, are 
\begin{subequations}
\bea
(1920\pi M^2)\, \langle T_t^t \rangle
  & = &  -40-20\xi-60 l(l+1)+80\xi\mu^2+20\xi l(l+1)
         +f(170+160\mu^2
         \nonumber \\
  &   &  \mbox{}
         +220 l(l+1)-480\mu^4-1334\xi-1840\xi\mu^2-720\xi\mu^4
         -30 l^2(l+1)^2
         \nonumber \\
  &   &  \mbox{}
         -360\xi\mu^2 l(l+1)-1690\xi l(l+1)-240\mu^2 l(l+1)-45\xi l^2(l+1)^2)
         \nonumber \\
  &   &  \mbox{}
         +\ln(1+\frac{3 l(l+1)+2}{12\mu^2})\left[
             240\mu^2+60 l(l+1)-480\xi\mu^2-120\xi l(l+1)
             \right. \nonumber \\
  &   &      \left. \mbox{}
             +f(480\mu^2+2880\mu^4
             -1920\xi\mu^2-8640\xi\mu^4+180 l^2(l+1)^2
             \right. \nonumber \\
  &   &      \left. \mbox{}
             +1440\mu^2 l(l+1)-120\xi l(l+1)
             -540\xi l^2(l+1)^2-4320\xi\mu^2 l(l+1))
         \right] \nonumber \\
  &   &  \mbox{}
         +\ln(1+\frac{ l(l+1)+1}{4\mu^2})\left[
             120\xi+480\xi\mu^2+120\xi l(l+1)
             +f(60-480\mu^2
             \right. \nonumber \\
  &   &      \left. \mbox{}
             -120 l(l+1)-2880\mu^4+480\xi+4800\xi\mu^2+11520\xi\mu^4
             +1200\xi l(l+1)
             \right. \nonumber \\
  &   &      \left. \mbox{}
             +5760\xi\mu^2 l(l+1)+720\xi l^2(l+1)^2
             -1440\mu^2 l(l+1)-180 l^2(l+1)^2)
         \right] \nonumber \\
  &   &  \mbox{}
         +\ln(1+\frac{ l(l+1)+2}{4\mu^2})\left[
             f(-1080\xi-2880\xi\mu^2-2880\xi\mu^4
             \right. \nonumber \\
  &   &      \left. \mbox{}
             -180\xi l^2(l+1)^2
             -1080\xi l(l+1)-1440\xi\mu^2 l(l+1))
         \right]
         + O(f^2) \, ,  
\eea
\bea
(1920\pi M^2)\, \langle T_r^r \rangle
  & = &  -40-20\xi-60 l(l+1)+80\xi\mu^2+20\xi l(l+1)
         +f(70+60 l(l+1)
         \nonumber \\
  &   &  \mbox{}
         -160\mu^4-498\xi-400\xi\mu^2-240\xi\mu^4
         -120\xi\mu^2 l(l+1)-80\mu^2 l(l+1)
         \nonumber \\
  &   &  \mbox{}
         -10 l^2(l+1)^2-510\xi l(l+1)-15\xi l^2(l+1)^2)
         \nonumber \\
  &   &  \mbox{}
         +\ln(1+\frac{3 l(l+1)+2}{12\mu^2})\left[
             240\mu^2+60 l(l+1)-480\xi\mu^2-120\xi l(l+1)
             \right. \nonumber \\
  &   &      \left. \mbox{}
             +f(480\mu^2+960\mu^4
             -1920\xi\mu^2-2880\xi\mu^4
             -360\xi l(l+1)
             \right. \nonumber \\
  &   &      \left. \mbox{}
             -180\xi l^2(l+1)^2
	     +480\mu^2 l(l+1)-1440\xi\mu^2 l(l+1)+60 l^2(l+1)^2)
         \right] \nonumber \\
  &   &  \mbox{}
         +\ln(1+\frac{ l(l+1)+1}{4\mu^2})\left[
             120\xi+480\xi\mu^2+120\xi l(l+1)
             +f(-60-480\mu^2
             \right. \nonumber \\
  &   &      \left. \mbox{}
             -120 l(l+1)-960\mu^4+480\xi+2880\xi\mu^2
             +3840\xi\mu^4+720\xi l(l+1)
             \right. \nonumber \\
  &   &      \left. \mbox{}
             -60 l^2(l+1)^2
             +240\xi l^2(l+1)^2-480\mu^2 l(l+1)+1920\xi\mu^2 l(l+1))
         \right] \nonumber \\
  &   &  \mbox{}
         +\ln(1+\frac{ l(l+1)+2}{4\mu^2})\left[
             f(-360\xi-960\xi\mu^2-960\xi\mu^4-360\xi l(l+1)
             \right. \nonumber \\
  &   &      \left. \mbox{}
             -60\xi l^2(l+1)^2-480\xi\mu^2 l(l+1))
         \right]
         + O(f^2) \, ,  
\eea
\bea
(192\pi M^2)\, P
  & = &  1-4\mu^2-4\xi- l(l+1)+16\xi\mu^2+4\xi l(l+1)
         \nonumber \\
  &   &  \mbox{}
         +\ln(1+\frac{ l(l+1)+1}{4\mu^2})
         [-6-24\mu^2+24\xi-6\, l(l+1)+96\xi\mu^2+24\,\xi\, l(l+1)]
         \nonumber \\
  &   &  \mbox{}
         +\ln(1+\frac{3 l(l+1)+2}{12\mu^2}) [24\mu^2-96\xi\mu^2-24\,\xi\, l(l+1)]
         + O(f) \, .
\eea
\end{subequations}
The stress tensor and pressure for the conformal field are retrieved 
by taking $m$, $\xi$, and $l$ to zero \footnote{An error was made in 
equation (4.16) of \cite{BaFaNiFrSuZe:01} which affected the finite 
terms in (4.20), (4.21), and (4.23) of that paper.  Equation (4.16) 
should read 
\ben
\langle T_a{}^b\rangle_0=O(f^2) \, . 
\een
As a consequence, in (4.20) the factor $17/240$ should be replaced 
by $1/24$, in (4.21) the factor $-83/960$ should be replaced by $-11/192$,
and in (4.23) the factor $-23/240$ should be replaced by $1/48$.  
The conclusions regarding the regularity of the field at the horizon 
are unchanged.  
Note also that in \cite{BaFaNiFrSuZe:01} $P$ was defined with an 
additional factor of $(4\pi r^2)^{-1}$.}. 

The challenge remains to extend these explicit formulae for the 
stress tensor and pressure outside of the immediate neighborhood 
of the horizon, to match onto the \hfa  
(\ref{Schw}).

\section{Conclusions}
\label{conclusions}

In this paper we have obtained an analytic approximation for the
effective action of a quantum scalar field in a general static
two-dimensional spacetime. The resulting expression shows 
clearly both the zero and the finite temperature contributions; 
these latter are competely missed in the construction based on 
functional integration of the conformal anomaly. 

In this elegant and rather simple way an expression for the stress 
tensor of the scalar field can be obtained and is shown to coincide 
with the one obtained by standard canonical construction under the WKB
approximation. The most remarkable aspect of this new method is
that the same procedure can be repeated for the full 4D case
\cite{FrSuZe:00,FrSuZe:02,Su:00b}, reproducing exactly the results 
of Anderson, Hiscock, and Samuel \cite{AnHiSa:95}. 

Unfortunately our effective action, like other analytical approaches, 
is unable to handle correctly the horizon region, where the
high-frequency approximation breaks down. This causes the
appearance of unphysical divergences in the stress tensor in the
Hartle-Hawking state. We have shown how a careful analysis of the
Green function near the horizon leads to the expected regular
result. 

Much effort is still needed to find an analytic expression for 
the stress tensor in black-hole spacetimes which correctly describes 
both the horizon and the asymptotic region. Only then can a serious 
analysis of nonperturbative backreaction effects in black-hole 
spacetimes start.


\section*{Acknowledgments}

The authors are grateful to Valeri Frolov and Andrei Zelnikov for helpful 
discussions.  PJS would like to thank the INFN and the Natural Sciences 
and Engineering Research Council of Canada for their financial support.  
This work has been funded by NSF grant PHY~00-99559 and its predecessor.  
The Center for Gravitational Wave Physics is supported by the NSF under 
co-operative agreement PHY~01-14375.


\appendix
\renewcommand{\theequation}{\thesection.\arabic{equation}}
\section{WKB Approximation for the $\bm{l=0}$ mode}\setcounter{equation}{0}
\label{WKBappendix}

Here we outline the calculation of the stress tensor of 
the two-dimensional s-wave field $\widehat{\varphi}_0$ satisfying 
(\ref{2dfieldeqn}) with $l=0$ in the Schwarzschild spacetime 
(\ref{metric}) using the WKB approximation with point-splitting 
regularization.  We show that the WKB approximation coincides 
exactly with the high-frequency approximation 
(\ref{SchwTtt}, \ref{SchwTrr}) for $l=0$.  
These calculations closely follow analogous ones performed 
in \cite{BaFaNiFrSuZe:01}; 
details on the method can be found there and in the seminal paper 
by Anderson, Hiscock and Samuel \cite{AnHiSa:95}. 

To take advantage of the work of \cite{BaFaNiFrSuZe:01,AnHiSa:95} 
it is convenient to consider not the two-dimensional Euclidean 
Green function $G$ of (\ref{fieldeqn}) but the rescaled quantity
\be\n{G_E}
G_E(t,r;t',r') \equiv r r' G(t,r;t',r') \, . 
\ee
Substitution into (\ref{fieldeqn}) shows that $G_E$ satisfies 
\be 
[\nabla^a(e^{-2\phi}\nabla_a) - m^2 e^{-2\phi}]\,G_E(x,x') 
  = -\frac{\delta^2(x-x')}{\sqrt{h}} \, , 
\ee 
where $h_a^{c'}$ is the bivector of parallel transport. 
The expressions (\ref{stressG}), (\ref{stressD}) for the unrenormalized 
point-split stress tensor become  
\bea  
\left<T_{ab}(x,x')\right>
  & = &  e^{-\phi(x)-\phi(x')}\left\{
             (\frac{1}{2}-\xi)(h_a^{c'}G_{E;c'b}+h_b^{c'} G_{E;a c'})
             +(2\xi-\frac{1}{2}) h_{a b} h^{c d'}G_{E;cd'}
	     \right. \nonumber \\ 
  &   &      \mbox{} 
             -\xi(G_{E;ab}+h_a^{c'}h_b^{d'} G_{E;c'd'}) 
             +\left.(2\xi-\frac{1}{2}) h_{a b} m^2 G_E \right\}  
	     \, .
\eea 

We begin with the WKB approximation at zero temperature, and then 
calculate finite-temperature effects.  We conclude by showing  
how to estimate the contribution of the zero-frequency mode at 
finite temperatures in flat space; this is useful for correcting 
the WKB and high-frequency approximations in asymptotically flat regions. 

\subsection{Zero Temperature}

At zero temperature, this Green function has the integral expansion  
\be 
G_E(x,x')
  =  \int_0^\infty \frac{d\omega}{4\pi} \,
     \cos{[\omega(t-t')]} \,p_\omega(r_<)q_\omega(r_>) \, , 
\ee 
where $p_\omega$ and $q_\omega$ obey the differential equation
\be
f\frac{d^2 S}{dr^2} +[\frac{2f}{r}+f']\frac{dS}{dr}
-(\frac{\omega^2}{f}+m^2)S=0 \, .
\ee 
One can write $p_\omega$ and $q_\omega$ in the WKB form 
\begin{subequations}\n{pq}
\bea
p_\omega 
  & = &  \frac{1}{r\sqrt{2W}}\,
         \exp{\left\{\int^r\!\!dr\,\frac{W(r)}{f}\right\}} 
         \, , \\   
q_\omega
  & = &  \frac{1}{r\sqrt{2W}}\,
         \exp{\left\{-\int^r\!\!dr\,\frac{W(r)}{f}\right\}} 
         \, , 
\eea
\end{subequations}
where the function $W$ satisfies the equation 
\be 
W^2=\omega^2 + m^2f + V
+\frac{f}{2W}\left[f\frac{d^2W}{dr^2} +\frac{df}{dr}\frac{dW}{dr}
-\frac{3f}{2W}(\frac{dW}{dr})^2\right]
\ee 
with 
\be
V=\frac{f}{r}\frac{df}{dr}  \, .
\ee 
This is solved iteratively starting with the zeroth-order solution 
\be\n{zeroth} 
W=\sqrt{\omega^2+m^2f} \, .
\ee 
In the following we omit terms coming from the conformally
invariant part ($\xi=0$, $m=0$), indicated with $\left<T_a{}^b\right>^0$, 
which have been treated in \cite{BaFaNiFrSuZe:01}. We have 
\begin{subequations}\n{T_WKB}
\bea 
\left< T_t{}^t \right>_{unren}
  & = &  \left<T_t{}^t\right>_{unren}^0 
         +\e^{-2\phi}\int_0^\infty \frac{d\omega}{4\pi}
	 \,\cos\(\omega{\epsilon}_{\tau}\) \left\{
           -\frac{1}{2}m^2A_1 +\xi [ c_0{\omega}^2 A_1 + c_1 A_2 \quad
	   \right. \nonumber \\ 
  &   &    \left. \vphantom{\frac{1}{1}} \mbox{}  
           + c_2 A_3 + c_3\( A_4 + A_5 \)+c_4 A_5 + c_5 A_1 ]
	 \right\}
         \nonumber \\
  &   &  \mbox{} 
         +i\xi\e^{-2\phi}\int_0^\infty \frac{d\omega}{4\pi}\,\omega\,
	 \sin{\(\omega{\epsilon}_{\tau}\)} [c_6 A_1 + c_7\(A_4 + A_5\)] 
         \, , \\  
\left<T_r{}^r\right>_{unren}
  & = &  \left<T_r{}^r\right>_{unren}^0
         +\e^{-2\phi}\int_0^\infty \frac{d\omega}{4\pi}
	 \,\cos\(\omega{\epsilon}_{\tau}\)\left\{ 
           -\frac{1}{2}m^2A_1 +\xi [ c_8\omega^2 A_1 + c_9 A_3 
	   \right. \nonumber \\ 
  &   &    \left. \vphantom{\frac{1}{1}}\mbox{} 
           + c_{10} \(A_4 + A_5\)]
	 \right\}
         +i\xi\e^{-2\phi}\int_0^\infty\frac{d\omega}{4\pi}\,\omega
	 \,\sin{\(\omega{\epsilon}_{\tau}\)}c_{11} A_1 \, ,   
\eea 
\end{subequations}
where $\epsilon=t-t'$ (for convenience the points are split in time 
only, so $r'=r$ ), the $c_i$ at leading orders are  
\bea\n{c} 
& & c_0=\frac{2}{f}\(1+\frac{M^2\epsilon^2}{r^4}\)\, , \quad 
c_1=2h^{rr'}\, , \quad 
c_2=O\(\epsilon^2\)\, , \nonumber \\ & & 
c_3=-\frac{f^{\prime}}{2}\, , \quad 
c_4=O\(\epsilon^2\),\ \ c_5=2 m^2, \nonumber\\& & 
c_6=-h^{tt'}h^{tr'}f^{\prime}\, , \quad 
c_7=2h^{rt'}\, , \quad  
c_8=-\frac{M^2\epsilon^2}{fr^4}\, , \nonumber \\  & & 
c_9=O\(\epsilon^2\)\, ,  \quad 
c_{10}=\(\frac{2f}{r}+\frac{f^\prime}{2}\)\, , \quad 
c_{11}=h^{rt'}\frac{f^{\prime}}{f}\, ,  
\eea 
and 
\bea\n{A} 
A_1 &=& p_\omega q_\omega\, ,\quad 
A_2=\frac{dp_\omega}{dr}\frac{dq_\omega}{dr}\, ,\quad 
A_3=\frac{1}{4}p_\omega q_\omega\, ,\nonumber \\  
A_4&=&q_\omega\frac{dp_\omega}{dr}\, ,\quad
A_5=p_\omega\frac{dq_\omega}{dr} \, . 
\eea 
The expansions for the bivector of parallel transport are  
\bea
h^{tt'} & = &  -\frac{1}{f}-\frac{f'^2}{8f}\epsilon^2+O(\epsilon^4) \, ,
               \nonumber \\ 
h^{tr'} & = &  -h^{rt'} = -\frac{f'}{2}\epsilon+O(\epsilon^3) \, , 
               \nonumber \\ 
h^{rr'} & = &  f+\frac{f'^2 f}{8}\epsilon^2 +O(\epsilon^4) \, .
\eea
Eventually one arrives at the following expression 
for $\left<T_a{}^b\right> $ in the zero temperature state: 
\begin{subequations}
\bea\label{UN} 
\left< T_t{}^t\right>_{unren}
  & = &  \left<T_t{}^t\right>_{unren}^0
         +\frac{m^2}{4\pi} L 
         +\frac{\xi}{2\pi f}\left\{ 
	   -\frac{4Mf}{r^3}-\frac{M^2}{r^4}
           +[\frac{2Mf}{r^3}-\frac{2f^2}{r^2}] L 
	 \right\} 
         \, , \qquad \\ 
\left<T_r{}^r\right>_{unren}
  & = &  \left<T_r{}^r\right>_{unren}^0
         +\frac{m^2}{4\pi} L 
         +\frac{\xi}{2\pi f}\left\{
           \frac{M^2}{r^4} 
           +[\frac{2Mf}{r^3}+\frac{4f^2}{r^2} ] L 
	 \right\} \, , 
\eea 
\end{subequations}
where $L=[\frac{1}{2}\ln{\(-\lambda^2\epsilon^2\)}+\gamma]$ 
and $\gamma$ is Euler's constant. To obtain the
renormalized stress tensor, one subtracts from the above
expression the renormalization counterterms obtained from the
DeWitt-Schwinger Green function,  
\bea 
G_{DS}
  & = &  \frac{e^{\phi(x) +\phi(x')}}{2\pi}\left\{ 
             -L'[ 1+\alpha\sigma +O(\sigma^2)]
             +\frac{a_1}{2m^2}+\beta\sigma +O(\sigma^2) 
	 \right\}\, , 
\eea
where 
\be
L' = \frac{1}{2} \ln{\frac{m^2\sigma}{2}} + \gamma \, , 
\ee
$a_1$ is the DeWitt-Schwinger coefficient, 
\be
a_1=\frac{1}{6}(R-6(\nabla\phi)^2+6\Box\phi) \, , 
\ee 
and 
\begin{subequations}
\bea
\alpha&=&\frac{m^2}{2}+\frac{R}{12}-\frac{a_1}{2}\, , \\
\beta &=& \frac{m^2}{2}-\frac{a_1}{4} \, .
\eea 
\end{subequations}
Recall that $R$ is the scalar curvature 
of the \twod metric $h_{ab}$; $\sigma$ is one-half the square of the
geodesic distance between the points $x$ and $x'$. Performing all the 
calculations, the counterterms read 
\begin{subequations}
\bea 
\left< T_t{}^t\right>_{DS}&=&\left<
T_t{}^t\right>^0_{DS} +\frac{m^2}{4\pi} L' +\frac{\xi}{2\pi f}\lc
\frac{2Mf}{r^3}L'-\frac{2f^2}{r^2}L' \rc +\frac{m^2}{8\pi} \, , 
\\ 
\left< T_r{}^r\right>_{DS} &=&\left<
T_r{}^r\right>^0_{DS}+\frac{m^2}{4\pi} L' +\frac{\xi}{2\pi f}\lc
-\frac{2Mf}{r^3}+ \frac{2Mf}{r^3}L'+\frac{4f^2}{r^2}L' \rc 
-\frac{m^2}{8\pi} \, .\label{DS} \quad
\eea
\end{subequations}
Subtracting these from the unrenormalized expressions (\ref{UN}) 
one obtains the renormalized stress tensor in the Boulware vacuum: 
\begin{subequations}
\bea
\left<B|T_t{}^t|B\right>_{ren} 
  & = &\left<B|T_t{}^t|B\right>_{ren}^0
  -\frac{m^2}{8\pi}\ln(\frac{m^2f}{4\lambda^2}) 
  + \frac{\xi}{2\pi f}\left\{
  -\frac{4M}{r^3}+\frac{7M^2}{r^4}
  \right. \nonumber \\
  & & \left. \mbox{} 
  +\frac{f}{r^2}\(1-\frac{3M}{r}\)\ln\(\frac{m^2 f}{4\lambda^2}\) \right\} 
  -\frac{m^2}{8\pi} \, ,\\
\left<B|T_r{}^r|B\right>_{ren} 
  & = & \left<B|T_r{}^r|B\right>_{ren}^0
  -\frac{m^2}{8\pi}\ln(\frac{m^2f}{4\lambda^2}) 
  + \frac{\xi}{2\pi f}\left\{\frac{2M}{r^3}-\frac{3M^2}{r^4}
  \right. \nonumber \\
  & & \left. \mbox{} 
  +\frac{f}{r^2}\(\frac{3M}{r}-2\)\ln\(\frac{m^2 f}{4\lambda^2}\) \right\} 
  +\frac{m^2}{8\pi}\, , 
\eea
\end{subequations}
where the conformal ($\xi=0$, $m=0$) contribution is \cite{BaFaNiFrSuZe:01}  
\begin{subequations}
\bea 
\left<B | T_t{}^t|B \right> 
  &=&\frac{1}{2\pi f}\left\{ \frac{1}{12}\frac{M^2}{r^4}
  -\frac{1}{6}\frac{fM}{r^3} 
  -\frac{f^2}{4r^2}\ln\(\frac{m^2f}{4\lambda^2}\) \right\} 
  \, , \\  
\left<B | T_r{}^r|B\right> 
  &=& \frac{1}{2\pi f}\left\{ -\frac{1}{12}\frac{M^2}{r^4}
  -\frac{1}{2}\frac{fM}{r^3} 
  +\frac{f^2}{4r^2}\ln\(\frac{m^2 f}{4\lambda^2}\)\right\}\, .
\eea 
\end{subequations}

\subsection{Finite-Temperature Corrections}

The thermal case is treated similarly.  Evaluating the sum over 
discrete frequencies using the Plana sum formula \cite{ChVi:78} one finds
that the stress tensor at finite temperature is obtained from the
zero-temperature one by making the substitution 
\be
\ln(\frac{m^2f}{4\lambda^2})\to \left\{ 2\gamma
+\ln(\frac{m^2\beta^2f}{16\pi^2} ) \right\} 
\ee 
and adding the traceless pure radiation term 
\be 
\left< T_t{}^t\right>_{rad}
  =  -\left<T_r{}^r\right>_{rad}
  =  -\frac{\pi}{6\beta^2 f} \, , 
\ee 
where $\beta=T^{-1}$ is the inverse temperature.

\subsection{Zero-Frequency Mode}

One can also use the WKB procedure to calculate the contribution of 
the $n=0$ mode at nonzero temperature in the asymptotically flat 
region, as mentioned in Section~\ref{srg}.  In this case the zeroth-order 
solution (\ref{zeroth}) for $W$ is sufficient,
\be
W = m\sqrt{f} \, .
\ee
Substituting into (\ref{pq}) and using (\ref{c}), (\ref{A}) one can 
evaluate the contribution of this mode to the stress tensor, (\ref{T_WKB}). 
Note that for a field at finite temperature, we should replace the integrals 
over continuous $\omega$ in (\ref{T_WKB}) by sums.  For the zero-frequency 
mode alone, it is sufficient to make the replacement  
\be
\int\!\frac{d\omega}{4\pi}\,F(\omega) \longrightarrow \frac{F(0)}{\beta}\, .
\ee
See, for example, the equations following (3.4) of \cite{BaFaNiFrSuZe:01}.
Evaluating (\ref{T_WKB}), we find that the only term in the stress tensor 
which is nonvanishing at spatial infinity is 
\be
\langle T_r{}^r \rangle_{n=0} = -\frac{m}{2\beta\sqrt{f}}\, .
\ee


\section{Corrected $\bm{G_n}$ near the Horizon}
\label{correctedG}
\setcounter{equation}0

The solutions to the differential equation (\ref{ODE}) for the
contribution $G_n$ to the \twod Green function for modes of small $n$
are
\begin{subequations}
\bea
G_0(r,r')
  & = &  L_<^{\frac{1}{2}} \, I_0\left(\frac{\alpha_0 L_<}{M}\right) \, \left(
             1 - A_0^2 \left(\frac{L_<}{4M}\right)^4
         \right)
     \nonumber \\
  &   &  \mbox{}
     \times L_>^{\frac{1}{2}} \, \left(
         K_0\left(\frac{\alpha_0 L_>}{M}\right)
             -\frac{1}{2}A^2_0\left(\frac{L_>}{4M}\right)^4\left[
             1-2\gamma-2\ln\left(\frac{\alpha_0L_>}{2M}\right)
         \right]
     \right) \label{soln0} \, , \quad \\
G_1(r,r')
  & = &  L_<^{\frac{1}{2}} \, I_1\left(\frac{\alpha_1 L_<}{M}\right) \, \left(
             1 - \frac{2}{3}A_1^2 \left(\frac{L_<}{4M}\right)^4
         \right)
     \nonumber \\
  &   &  \mbox{}
     \times L_>^{\frac{1}{2}} \,
         K_1\left(\frac{\alpha_1 L_>}{M}\right) \, \left(
             1-2 A^2_1\left(\frac{L_>}{4M}\right)^4
     \right) \label{soln1} \, ,  \\
G_2(r,r')
  & = &  L_<^{\frac{1}{2}} \, I_2\left(\frac{\alpha_2 L_<}{M}\right) \, \left(
             1 - \frac{1}{2}A_2^2 \left(\frac{L_<}{4M}\right)^4
         \right)
     \nonumber \\
  &   &  \mbox{}
     \times L_>^{\frac{1}{2}} \, \left(
         K_2\left(\frac{\alpha_2 L_>}{M}\right)
             -\frac{1}{2\alpha_2^2}A^2_2\left(\frac{L_>}{4M}\right)^2\left[
             \gamma+\ln\left(\frac{\alpha_2L_>}{2M}\right)
         \right]
     \right) \label{soln2} \, ,  \\
G_3(r,r')
  & = &  L_<^{\frac{1}{2}} \, I_3\left(\frac{\alpha_3 L_<}{M}\right) \, \left(
             1 - \frac{2}{5}A_3^2 \left(\frac{L_<}{4M}\right)^4
         \right)
     \nonumber \\
  &   &  \mbox{}
     \times L_>^{\frac{1}{2}} \,
         K_3\left(\frac{\alpha_3 L_>}{M}\right) \, \left(
             1+2 A^2_3\left(\frac{L_>}{4M}\right)^4
     \right) \label{soln3} \, .
\eea
\end{subequations}
Here $L_>$ ($L_<$) is the physical distance from the event horizon
to the greater (lesser) of $r$, $r'$:
\be
L(r) = 2M\left[\,
           \sqrt{\frac{r}{2M}\left(\frac{r}{2M}-1\right)}
           +\ln{\left(\sqrt{\frac{r}{2M}}+\sqrt{\frac{r}{2M}-1}\right)}
       \,\right] \, .
\ee
The expansions of each of these functions for small $f$, $f'$
is significant up to and including the third term, and satisfies
the differential equation (\ref{ODE}) for the Green function to
the order required for obtaining the stress tensor to
$O(L^{\frac{1}{2}}\ln{L})=O(f\ln{f})$ inclusive.


\end{document}